# High-order tensor flow processing using integrated photonic circuits

Shaofu Xu[1]†, Jing Wang[1]†, Sicheng Yi[1], and Weiwen Zou[1]*

[1]State Key Laboratory of Advanced Optical Communication Systems and Networks, Intelligent Microwave Lightwave Integration Innovation Center (imLic), Department of Electronic Engineering, Shanghai Jiao Tong University; Shanghai, 200240, China

*Corresponding author. Email: wzou@sjtu.edu.cn

† These authors contributed equally to this work

*Abstract:* **Tensor analytics lays mathematical basis for the prosperous promotion of multiway signal processing. To increase computing throughput, mainstream processors transform tensor convolutions to matrix multiplications to enhance parallelism of computing. However, such order-reducing transformation produces data duplicates and consumes additional memory. Here, we demonstrate an integrated photonic tensor flow processor without tensor-matrix transformation, which outputs the convolved tensor as the input tensor 'flows' through the processor. The hybrid manipulation of optical dimensions of wavelength, time, and space enables the direct representation and processing of high-order tensors in optical domain. In the proof-of-concept experiment, processing of multi-channel images and videos is accomplished at the frequency of 20 GHz. A convolutional neural network is demonstrated on the processor, which achieves an accuracy of 97.9% on action recognition.**

Stacking data of multiple dimensions to form a tensor provides us the opportunity to discover the intrinsic structural features hidden in the data [1], which is invisible from two-way (matrix) data analysis. For example, multiway representation of electroencephalogram (EEG) data is the natural and effective way of neuroscience data processing [2] and the tensor stacked across time, space, and spectrum is beneficial to detect features in electromagnetic waveforms [3]. Since tensor matches high-dimensional nature of the world, the concept of multiway analytics gives rise to extensive signal processing approaches in fields including life science [2, 4], radar [5, 6], data mining [7, 8], and machine learning [9-11]. Among the basic operations for tensors, convolution is effective to extract structural features from data. Targeted features are filtered out as the convolutional kernel traverses the tensor. As an epitome, convolutional neural network, which plays a fundamental role of modern artificial intelligence (AI), is designed under the concept of multi-channel tensor processing [12, 13].

Given the fact that tensor processing, especially in the AI field, is consuming an increasing portion of computing resources, high-throughput and energy-efficient processors are eagerly pursued [14]. Straightforward algorithm of computing a tensor convolution contains multiple nested loops, not compatible with high-throughput paralleled processing. Therefore, generalized matrix multiplication (GeMM) approach is widely used in state-of-the-art high-performance computing (HPC) hardwares [15]. For example, in the Tensor Core of Nvidia Ampere architecture [16], the CUBE core of Huawei Davinci architecture [17], the systolic array of Google TPU architecture [18], and the cross-bar array of memristor architecture [19, 20], high-order tensor convolutions are transformed to two-dimensional matrix multiplications so that the depth of loops is lowered and paralleled computational cores can work simultaneously to enhance throughput. However, during the GeMM transformation, the input tensor should be duplicated and shifted for many times (related with the kernel size) to form an input matrix, which significantly increases memory use and additional memory access.

Besides electronic HPC processors, photonics is recently demonstrated as a promising candidate to build high-performance matrix processors. By designing the photonic circuit as linear transformation functions, matrix multiplications can be accomplished as the light flies through the circuit [21-24]. The broadband spectrum of photonic circuits boosts the clock frequency to tens of Gigahertz ($10^9$ Hz) [25-27]. Consequently, photonic circuits are demonstrated as superior GeMM processors with high throughput and energy efficiency [28, 29]. In fact, another advantage of photonics compared with electronics is that the available degrees of freedom of light is rich. For example, wavelengths [22, 26, 29], guiding modes [30], time [31], and space [21, 23, 24] are successfully investigated to carry out linear transformations. If we take a hybrid use of such degrees of freedom of light, direct representation of high-order tensors can be feasible [32, 33], so that photonic circuits can process tensors directly instead of via the GeMM.

Here, we demonstrate an integrated photonic tensor flow processer (PTFP) which directly processes high-order tensors without transformation. Namely, tensor convolution is completed as the input tensor 'flows' through the photonic circuit. This is achieved by the hybrid manipulation of optical degrees of freedom of wavelengths, time, and space. During the photonic processing, there is no external memory access. Kernel weights are implemented inside the microring resonators (MRRs) of the PTFP and data registering is accomplished by the embedded optical delay structure. In a proof-of-concept experiment, we implemented a silicon-based integrated photonic chip to conduct the key parts of the PTFP. Empowered by the broadband capability of light, the photonic chip works at the speed of 20 GHz and is capable to achieve a compute density surpassing trillions of operations per second per square millimeter. By reconfiguring the parameters of MRRs to change kernel weights, tensor (including multi-channel images and video) processing is experimentally demonstrated. A CNN is trained to validate the PTFP chip. Classification accuracy of 97.9 % on the KTH dataset (video action recognition) [34] is achieved at the inference phase.

**Principle**

Basic principles of the GeMM and the PTFP are compared in Fig. 1(**A**). The dimensionality of the input tensor is denoted as [$D_{data}$, $C_{in}$], where $D_{data}$ is the size of data in a single input channel (e.g. $D_{data}$ represents [*Width*, *Height*] when the input is an image) and $C_{in}$ denotes the number of input channels. Different from conventional convolutional operation, tensor convolution with multiple input channels should yield multiple output channels. [$D_{data}$, $C_{out}$] denotes the dimensionality of the output tensor. Each output channel is obtained by summing all convolved results from every input channel. Therefore, the dimensionality of a complete kernel of tensor convolution is denoted as [$D_{kernel}$, $C_{in}$, $C_{out}$], where $D_{kernel}$ is the size of a single convolution. In order to compute tensor convolution via matrix multiplication, shown by the 'GeMM' part of Fig. 1(**A**), GeMM firstly transforms the input tensor to an input matrix with the dimensionality of [$D_{data}$, $D_{kernel} \times C_{in}$], where data volume is augmented by $D_{kernel}$ times. The additional data is generated by duplicating and shifting the original data, occupying more memories and taking more memory accesses. The kernel tensor is reshaped to two-dimensional matrix and then the output matrix is obtained by matrix multiplication. In the process of the PTFP (shown in 'Flow' part), the input tensor is not transformed. Different input channels are carried by different optical wavelengths. Data in a single channel is encoded onto time steps of an optical sequence. Inside the PTFP, each input channel is connected with each output channel through a convolutional operation (a line in the figure). A convolutional operation is essentially a finite impulse response (FIR) filter; therefore, we can implement such filters by imposing delaying, weighting, and summation to the input temporal sequence. The number of delays equals to the size of kernel, $D_{kernel}$. That means the additional memory required by GeMM is equivalently accomplished with optical delay structure in the PTFP. Given that the input sequences are carried on different wavelengths, the convolved sequences are combined together to yield an output channel with wavelength division multiplexing (WDM). Other output channels are similarly yielded by spatially duplicating the same structure but configuring different kernel weights.

**Results**

To demonstrate the PTFP concept, we design and fabricate a PTFP chip whose schematic is illustrated in Fig. 1(**B**). Input optical sequences of different wavelengths are firstly combined with a WDM. Then, directional couplers and delay lines are deployed to provide the dimension of time, $D_t$. In each time dimension, optical sequences are further split to provide the dimension of space, $D_s$. In a specific time and space dimension, a weighting bank with $D_w$ MRRs are exploited. A single MRR in a weighting bank controls the transmission rate of a specific wavelength. By shifting the resonance wavelength of MRRs, weights of input wavelengths can be reconfigured. Via these $D_w \times D_t \times D_s$ copies of MRRs, multiplications involved in a complete convolutional kernel is accomplished. After weighting, photodetectors (PDs) convert the total optical power of all wavelengths to electrical signals, performing summation across different input channels. And the electrical power combiners (EPC) perform electrical summations of signals across different delays. Since operations on the chip is linear, two steps of summations are commutative. Every output sequence of the EPC corresponds to an output channel in Fig. 1(**A**).

Figure 2(**A**) shows the photograph of the packaged PTFP chip, which is fabricated with standard Silicon-on-Insulator (SOI) integration process. As a proof-of-concept, we implement the key components of the PTFP onto the chip, including WDM, optical delays, and weighting banks. Optical signals enter and leave the chip through the waveguide-fiber edge coupler array. The fabricated chip holds the dimensions of [$D_w$=4, $D_t$=3, $D_s$=1]. Given the fact that the expansion of space dimension is duplicating the same structure for multiple times, successful validation of a chip with $D_s$ =1 provides strong evidence for additional space dimension. Figure 2(**B**) depicts the layout of the fabricated PTFP chip, comprising a four-way WDM, two cascaded ODLs, and three weighting banks with four MRRs inside each. The WDM shown in Fig. 2(**C**) is designed with the asymmetric Mach-Zehnder interferometer structure. Figure 2(**D**) presents the transmission rate measurement of the WDM, showing 2-nm channel spacing and <1.2 dB channel flatness within a free spectral range (FSR). In the experiment, we choose four wavelengths locating at 1550.8 nm, 1552.8 nm, 1554.8 nm, and 1556.8 nm to ensure that all operating wavelengths are within the flat band of the WDM. Figures 2(**E**) and 2(**F**) illustrate the photograph and characterization result of a weighting bank. By increasing the voltage on the MRR, the resonating wavelength is red-shifted. Since the operating wavelengths are fixed, the variation of the MRR transmission rate performs as a weighting factor to the specific wavelength. Figure 2(**G**) provides the normalized weights of every MRR with the variation of applied voltages.

With this weight-voltage mapping, we can configure the applied voltages to represent convolutional kernels.

To validate the tensor processing capability of the PTFP chip, we carry out an experiment with channel-stacked images as the input tensor. Fig. 3(**A**) illustrates the conceptual experimental setup. The PTFP chip accepts four input signals with different wavelengths. Each signal represents an input channel and a single channel is an image of the channel-stacked images. These images are firstly reshaped to a row vector row by row; thus, they can be encoded onto optical intensities via temporal modulation. Four-way signals are generated with the symbol rate of 20 Gbaud/s, also known as clock frequency of 20 GHz. Since the optical intensities of different wavelengths are summed up in the PD, it is necessary to carry out input synchronization to avoid symbol misalignment. Similarly, the output signals with different optical delays should be also synchronized since they are summed up in the EPC. We deploy tunable delay lines before and after the optical ports of the PTFP chip for synchronization. Fig. 3(**B**) shows the result of output synchronization. In this measurement, only one input channel is adopted, so the output waveform should be identical except for delay. We observe that, after synchronization, the delay difference of every output waveform is 50 ps, corresponding to the symbol rate of 20-Gbaud/s. Using one input channel, we can conduct 1×3 convolutions by applying weights on the MRRs. Fig. 3(**C**) is an example of the convolved waveform. The applied weights are [-1, 0, 1]. From the zoom-in plot, we observe that the experimental results are close to the theoretically calculated samples, verifying the correctness of conducting one-dimensional convolution.

As we have multiple channels for input, we can realize a 3×3 kernel by exploiting the property of tensor processing. We configure the dimensionality of input tensor as [$D_{data}$, 3] and that of the kernel tensor as [1×3, 3, 1]. In principle, such configuration represents carrying out three individual one-dimensional convolutions for three input channels and adding them together. By setting the input channels to be the same image with row shifting, the convolution result is equivalent to a 3×3 two-dimensional convolution. Figs. 3(**D**)-3(**G**) depict the convolutional results with several typical kernels. The horizontal Sobel kernel extracts gray scale variations along the horizontal direction, so the convolved image is composed with vertical edges. Similarly, the vertical Sobel kernel can extract horizontal edges of the image. A kernel with nine same weights can blur the image. When a Sobel kernel is superposed with an identical kernel, the image can be sharpened and the edge contrast is increased. The experimental results verify the capability of the PTFP chip to conduct tensor convolution.

Based on the successful validation of the PTFP chip's capability of tensor processing, we move forward to implement a CNN to recognize human actions in the KTH dataset. Fig. 4(**A**) gives the structure of the built CNN with two convolutional layers, a recurrent layer, and a fully connected layer. We generate 4998 video segments from the KTH dataset. 3998 segments out of them are used as trainset and the left 1000 segments are used as testset. The parameters of CNN are trained firstly on a computer and the PTFP chip is used in the inference phase. Five frames of video are input into the neural network as the input tensor. Similar to the experiment of image convolution, each input frame is reshaped to a row vector for the temporal modulation. For the first convolutional layer, the adopted kernel size is [1×3×3, 1, 4]. Given that the fabricated chip is smaller than the kernel size, the kernel is decomposed to small parts and calculated by recalling the PTFP chip for multiple times. The PTFP chip calculates a kernel of [1×3, 3, 1] for each time of recalling and accomplish the complete kernel for 4 times of recalling. The same decomposition method is used for calculating the second convolutional layer with kernel size of [1×3×3, 4, 8]. Figs. 4(**B**) and 4(**C**) display several experimental results of the first convolutional layer and the second convolutional layer, respectively. We observe that, the convolved frames output by the PTFP chip are consistent with that of a digital computer, except for some experimental noise. These two convolutional layers extracts frame features that contribute to action recognition. By finishing the following recurrent layer and fully connected layer in an auxiliary computing device, recognition result is obtained. The diffusion matrix with five categories of human actions ('boxing', 'handwaving', 'handclapping', 'walking', and 'running') is shown in Figs. 4(**D**) and a reference is offered in Fig. 4(**E**). Ninety-six video segments randomly selected from the testset are recognized. Numbers on the diagonal line counts correct recognitions. It is shown that the recognition accuracy of the PTFP chip is 94/96=97.9% and that of a digital computer is 95/96=98.9%. Recognition result confirms that the PTFP chip accomplishes tensor convolution successfully. We carry out simulations to reveal how the noise affects the recognition accuracy. Consistent with intuition, the accuracy tends to decrease with large noise amplitude (see Fig. 4(**F**)). The standard deviation of the experimental noise is around 0.1 and the achieved accuracy is higher than the situation with Gaussian noise at the same level ($\sigma_{noise}$ =0.1).

**Discussion**

In our experiment, the PTFP chip is operated at the speed of 20 Gbaud/s, corresponding to a throughput of 480 GOP/s. Given the footprint of the on-chip devices, the computing density of the PTFP chip is 588 GOP/s/mm$^2$. It is comparable with that of Nvidia A100 GPU (755 GOP/s/mm$^2$ for int-8) fabricated with 7-nm CMOS [16]. For a PTFP chip with larger scale, the computing density is capable to surpass 1 TOP/s/mm$^2$. Since the ODLs play a key role in the tensor processing, insertion loss and footprint of them are determinant to the signal-to-noise ratio, throughput and computing density of the PTFP chip. Given that the length of ODLs is inversely proportional to the clock frequency, advanced electrooptic modulators [35-37], PDs [38], and electrooptic packaging technologies [39, 40] are beneficial to shortening on-chip ODLs. Moreover, recent progress on ultra-low-loss silicon nitride waveguides [41] also offers the opportunity to implement complicated optical delay manipulations. Although the demonstrated kernel size in the proof-of-concept experiment is relatively small, it is straightforward to increase the kernel size. Using arrayed waveguide grating or MRRs as the WDM can increase wavelength dimension; cascading more ODLs is feasible to achieve higher time dimensions; and simply duplication the same structure can provide additional space dimensions. The only difference of a PTFP chip with larger kernel size is the additional introduction of waveguide crossing (virtual breaking points in Fig. 1(**B**)). Recent works show that insertion loss around 0.04 dB/crossing is obtainable [42, 43], implying that the influence of waveguide crossing can be minor. Although the validation of the PTFP concept is successful with the fabricated chip, there is large space to improve the integrated photonic devices for better performance. For example, the insertion loss of the WDM and ODLs should be lowered; the power flatness of the directional coupler should be promoted; tuning efficiency of MRR should be improved. With accessible silicon-photonic technologies and proper device design, these refinements are realizable so that the performance (e.g. signal-to-noise ratio and energy efficiency) of the chip can be greatly enhanced.

**Conclusion**

We propose and experimentally demonstrate an integrated photonic tensor flow processor which is, compared with current mainstream processors, capable to process high-order tensor convolutions without transformation and additional memory use. Wavelength dimension carries different channels of the input tensor and space dimension represents different channels of the output tensor. Between the input and the output, optical time delay, weighting, and summation perform convolutional operations. The hybrid manipulation of optical dimensions of wavelength, time, and space offers us the opportunity to process tensor in a 'flow' or 'non-stop' fashion. The PTFP chip is fabricated for the proof-of-concept experiment. Tensor processing with a four-order kernel tensor is demonstrated. Further, a CNN is built and the video action recognition task is performed with high accuracy of 97.9%, so that tensor processing capability of the PTFP chip is experimentally verified. Given the fact that a major performance bottleneck of state-of-the-art currently mainstream processors is the speed of memory access [44], the concept of tensor flow processing may become an effective way for future high-performance processors. Moreover, photonic ultrafast clock frequency not only makes tensor flow processing to be realizable but also boosts the throughput of photonic chips significantly. Therefore, the proposed PTFP concept is able to promote the advances of compute-intense applications such as video processing, high-resolution surveillance, autonomous driving, and Internet of Things.

**Acknowledgments:** The authors acknowledge Prof. Qunbi Zhuge for the usage of arbitrary waveform generator (Keysight M8194A).

**Funding:**

National Key Research and Development Program of China, Program No. 2019YFB2203700

National Natural Science Foundation of China, Grant No. 61822508

**Author contributions:**

Supervision: WZ

Funding acquisition: WZ

Conceptualization: SX, WZ

Methodology: SX, JW, SY

Investigation: SX, JW, SY

Visualization: SX, JW

Project administration: WZ

Writing – original draft: SX, JW

Writing – review & editing: SX, JW, WZ

**Competing interests:** Authors declare that they have no competing interests.

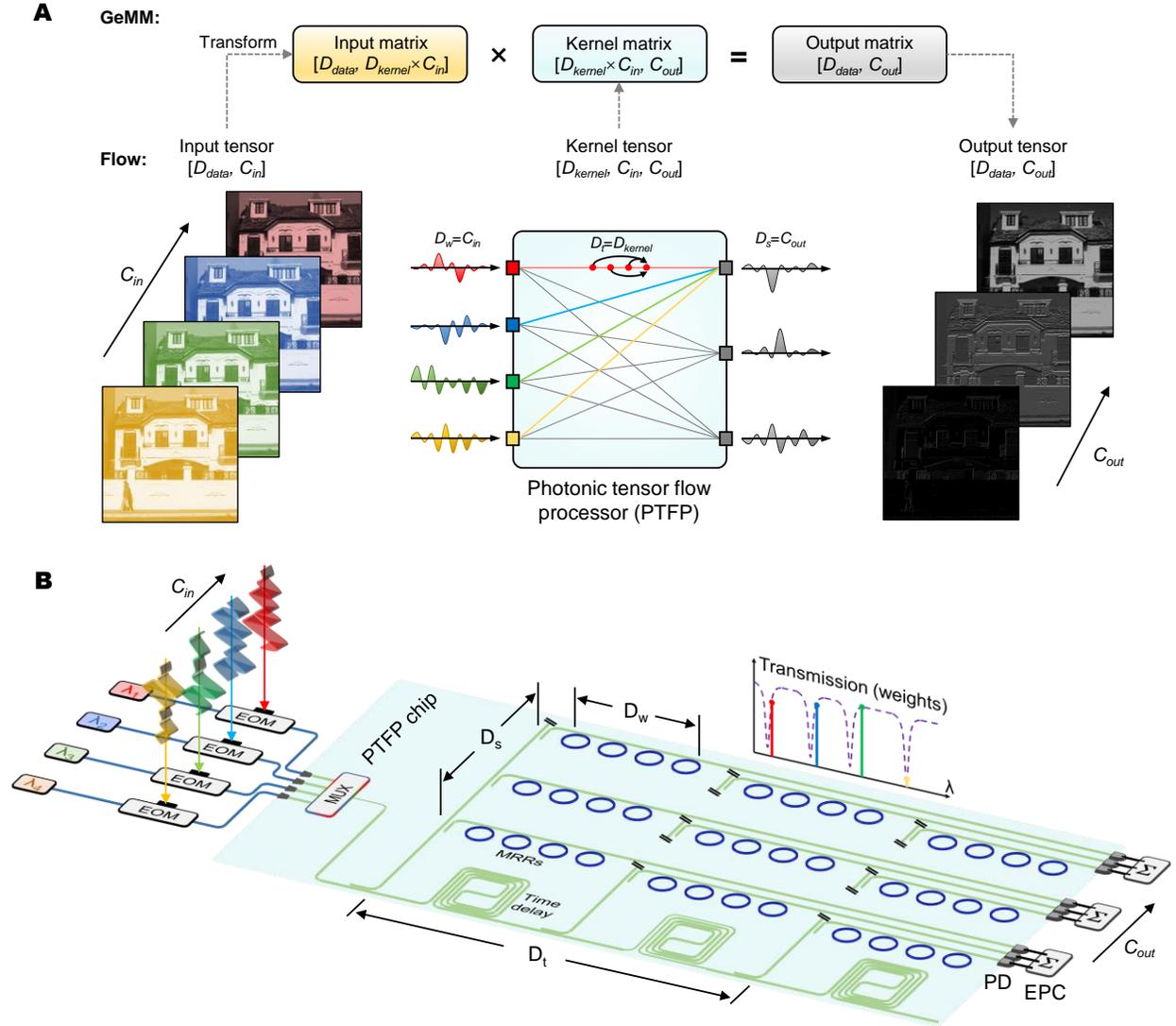

**Fig. 1. Basic principles of the PTFP.** (**A**) Principles of the conventional GeMM and the PTFP. Before matrix multiplication, the GeMM transforms the input tensor ($[D_{data}, C_{in}]$) to the input matrix ($[D_{data}, D_{kernel} \times C_{in}]$). The input data is reshaped to a column vector so that $D_{data}$ is one-dimensional. In the PTFP approach (marked with 'Flow'), the number of wavelength dimension ($D_w$) equals to the number of input channels. A photograph with different colors represents different input channels (the photograph was taken by an author). Every input channel is reshaped to a row vector and encoded onto the optical carrier via temporal intensity modulation. Inside the PTFP, every input channel is connected to every output channel with a line representing a convolutional operation. Inside a line, delaying and weighting are accomplished. Colored lines are used for highlighting the connections for one output channel, other output channels can be realized by spatially duplicating the same structure. (**B**) Conceptual schematic of the PTFP chip. EOM, electro-optic modulator; MUX, wavelength multiplexer; ODL, optical delay line. Directional couplers are depicted for light splitting. Crossing waveguides are virtually broken for succinctness of the graph.

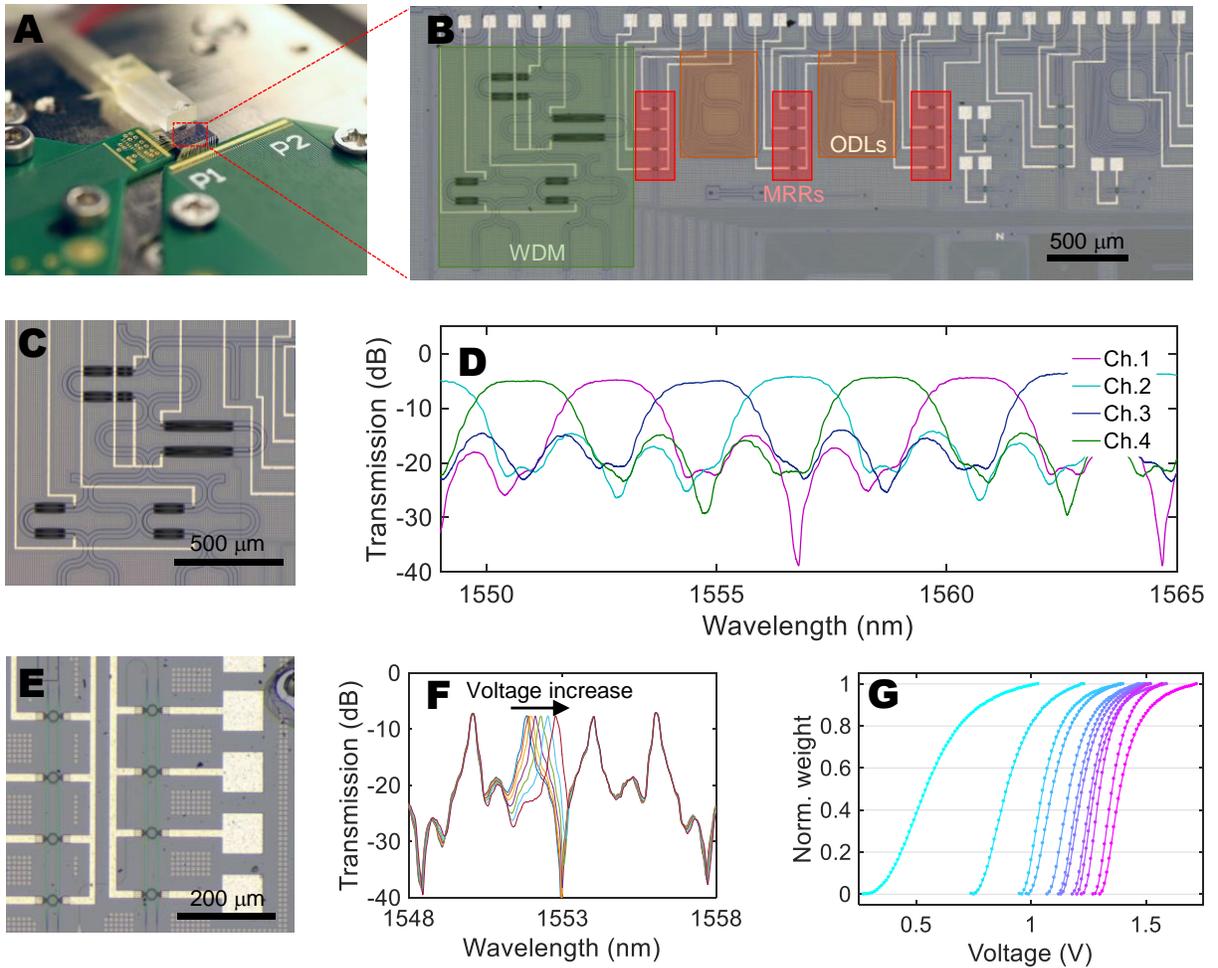

**Fig. 2. Chip fabrication and characterization.** (**A**) Photograph of the packaged PTFP chip. Optical signals enter and leave the chip via an edge-coupled fiber array. (**B**) Layout of the PTFP chip. Four wavelengths are combined in the WDM. Two optical delay lines (ODLs) are deployed to provide three temporal dimensions. Before and after each ODL, weighting banks with four MRRs in each are implemented. (**C**) Photograph of the WDM. (**D**) Transmission spectra of the WDM. (**E**) Regional photograph of the MRR array. (**F**) Transmission spectrum of the MRR array. Different voltages (0 to 1400 mV with 200 mV/step) are applied on the second MRR. Similar result can be obtained when voltage is applied on other MRRs. (**G**) Transmission rate of all twelve MRRs on the chip under voltage tuning. These curves represent weight-voltage mappings after normalization.

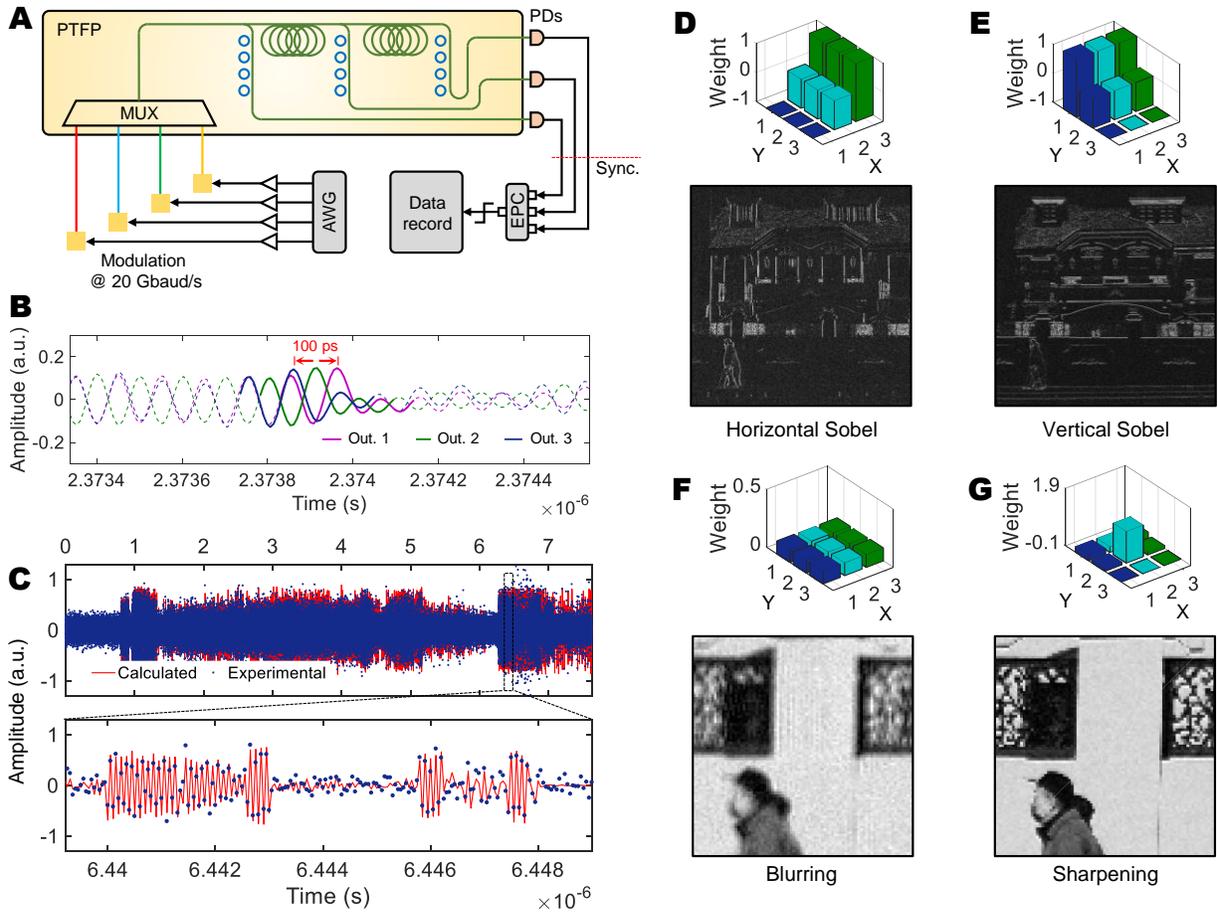

**Fig. 3 Experimental results of tensor convolution.** (**A**)Conceptional experimental setup of tensor convolution with the PTFP chip. MUX, wavelength multiplexer; AWG, arbitrary waveform generator. The generated signals are amplified and modulated on the optical carrier. After the tensor convolution, digital data is recorded at the output of the EPC. (**B**) Output synchronization. In the synchronization, only one waveform is used to input one signal. The waveforms of different output ports are identical with different delays. We highlight an identical segment of these waveforms with thicker linewidth. (**C**) Output samples of tensor convolution. A zoom-in plot is given for details. (**D**-**G**) Convolutional results with different applied kernels of horizontal Sobel, vertical Sobel, blurring, and sharpening, respectively. Weights of the kernels are provided by the bar charts. Subfigures of **F** and **G** show a bottom-left patch of the original image for better observation.

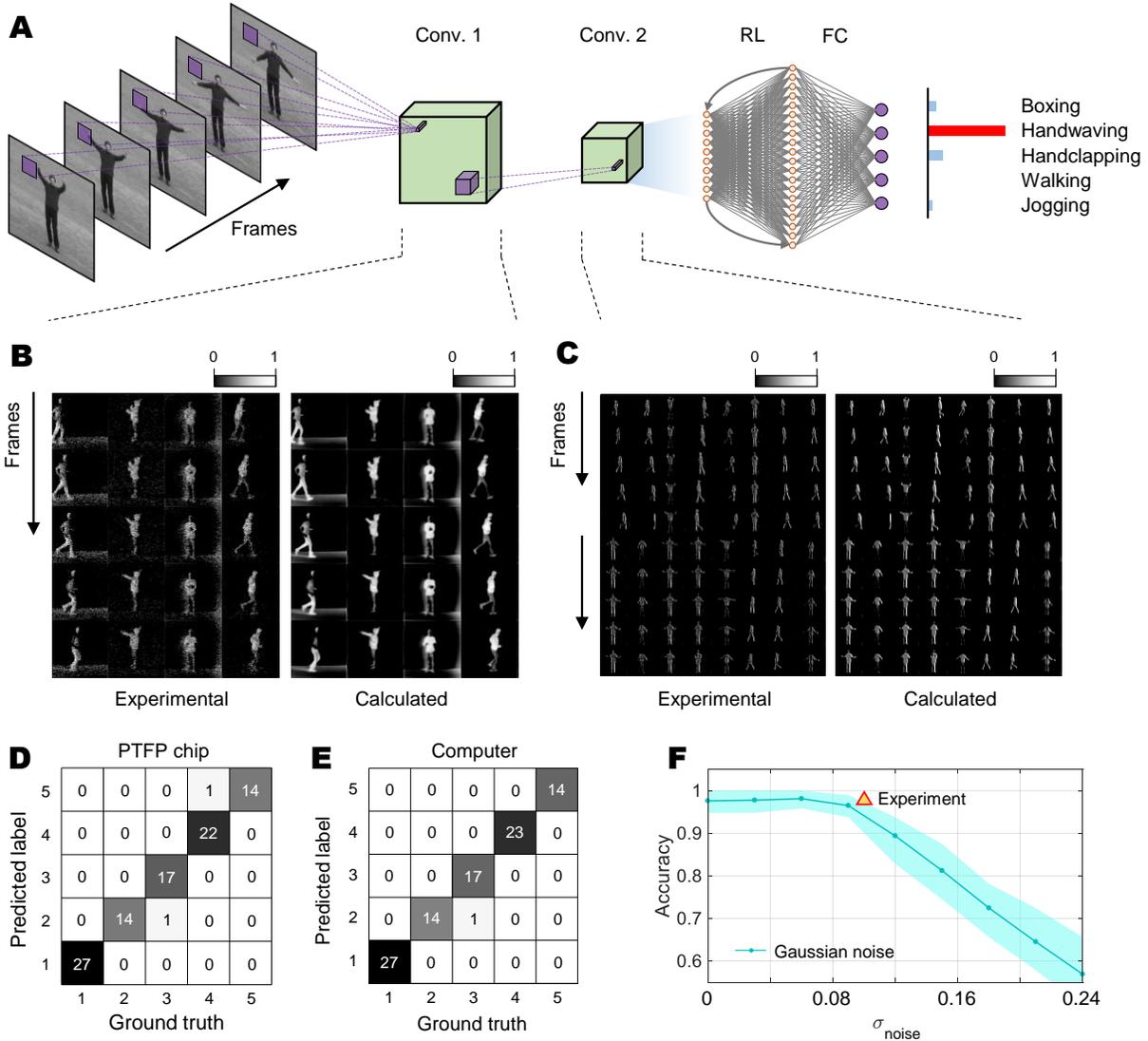

**Fig. 4 Experimental results of video action recognition of the KTH dataset.** (**A**) The adopted neural network model. Input data is a segment of video with 5 frames. Neural network is composed with two convolutional layers (Conv. 1 and Conv. 2), a recurrent layer (RL), and a fully connected layer (FC). The linear part of convolutional layers is computed by the PTFP chip. (**B**) and (**C**) Convolutional results of the PTFP chip of Conv. 1 and Conv. 2, respectively. Subplots from top to bottom display the convolved images of different frames. From left to right, several convolved video segments are displayed. For reference, computer-calculated results are provided aside. (**D**) and (**E**), Diffusion matrices of recognition of the PTFP chip and that of a digital computer, respectively. Numbers on the diagonal line record correct prediction. (**F**) Simulated accuracy of the neural network with different standard deviations ($\sigma_{noise}$) of additive Gaussian noise. The solid curve represents the average recognition accuracy and the shading indicates the 90% confidence interval. Yellow triangle marks the experimental accuracy of the PTFP chip.